\font\sixrm=cmr6
\font\sixi=cmmi6
\font\sixsy=cmsy6

\font\sevenrm=cmr7
\font\seveni=cmmi7
\font\sevensy=cmsy7

\font\tenrm=cmr10
\font\teni=cmmi10
\font\tensy=cmsy10
\font\tenit=cmti10
\font\tensl=cmsl10
\font\tenbf=cmbx10
\font\tentt=cmtt10

\font\twelverm=cmr12
\font\twelvei=cmmi12
\font\twelvesy=cmsy10 at 12pt
\font\twelveit=cmti12
\font\twelvesl=cmsl12
\font\twelvebf=cmbx12
\font\twelvett=cmtt12

\def\twelvepoint{%
\def\rm{\fam0\twelverm}%
\def\it{\fam\itfam\twelveit}%
\def\sl{\fam\slfam\twelvesl}%
\def\bf{\fam\bffam\twelvebf}%
\def\tt{\fam\ttfam\twelvett}%
\def\cal{\twelvesy}%
 \textfont0=\twelverm
  \scriptfont0=\sevenrm
  \scriptscriptfont0=\sixrm
 \textfont1=\twelvei
  \scriptfont1=\seveni
  \scriptscriptfont1=\sixi
 \textfont2=\twelvesy
  \scriptfont2=\sevensy
  \scriptscriptfont2=\sixsy
 \textfont3=\tenex
  \scriptfont3=\tenex
  \scriptscriptfont3=\tenex
 \textfont\itfam=\twelveit
 \textfont\slfam=\twelvesl
 \textfont\bffam=\twelvebf
 \textfont\ttfam=\twelvett
 \baselineskip=15pt
}
\def\tenpoint{%
\def\rm{\fam0\tenrm}%
\def\it{\fam\itfam\tenit}%
\def\sl{\fam\slfam\tensl}%
\def\bf{\fam\bffam\tenbf}%
\def\tt{\fam\ttfam\tentt}%
\def\cal{\tensy}%
 \textfont0=\tenrm
  \scriptfont0=\sevenrm
  \scriptscriptfont0=\sixrm
 \textfont1=\teni
  \scriptfont1=\seveni
  \scriptscriptfont1=\sixi
 \textfont2=\tensy
  \scriptfont2=\sevensy
  \scriptscriptfont2=\sixsy
 \textfont3=\tenex
  \scriptfont3=\tenex
  \scriptscriptfont3=\tenex
 \textfont\itfam=\tenit
 \textfont\slfam=\tensl
 \textfont\bffam=\tenbf
 \textfont\ttfam=\tentt
 \baselineskip=12pt
}

\font\authorrm=cmr12 scaled \magstep1

\font\titlebx=cmbx12 scaled \magstep2

% PAGE MAKEUP PARAMETERS
% ----------------------------------------------------------------

\hsize     = 148mm
\vsize     = 236mm
\hoffset   =    5mm
\voffset   =    4mm
\topskip   =  19pt
\parskip   =   0pt
\parindent =   0pt

% NEW COMMANDS
% ----------------------------------------------------------------
\newskip\one
\one=15pt

\newcount\LastMac
  % null element
\def\Skipe{1}  % SkipToFirstLine
    % text
    % heading A
    % heading B
    % heading C

\def\SkipToFirstLine{% move to start of text proper
 \LastMac=\Skipe
 \dimen255=150pt
 \advance\dimen255 by -\pagetotal
 \vskip\dimen255
}

\def\Raggedright{%
 \rightskip=0pt plus \hsize
 \spaceskip=.3333em
 \xspaceskip=.5em
}

\def\Fullout{% justify all lines
 \rightskip=0pt
 \spaceskip=0pt
 \xspaceskip=0pt
}

% DESIGN ELEMENTS
% ----------------------------------------------------------------

\def\ct#1\par{% chapter title
% \One
 %\Raggedright
 \titlebx\baselineskip=22pt
 #1
\vskip0.8truecm
}

\def\ca#1\par{% chapter author
% \One
 %\Raggedright
 \authorrm\baselineskip=18pt
 #1
\vskip0.8truecm
}

\def\aa#1\par{% author affiliation
% \One
 %\Raggedright
 \twelveit\baselineskip=15pt
 #1
\vskip0.5truecm
}

\def\ha#1\par{% heading A
% \ifnum\LastMac=\Skipe \else \One\fi
 %\LastMac=\Hae
 \Raggedright
\vskip15pt
 \twelvebf\baselineskip=15pt
 #1

\noindent
}

\def\hb#1\par{% heading B
% \LastMac=\Hbe
 %\One
 %\Raggedright
\vskip15pt
 \twelveit\baselineskip=15pt
 #1

\noindent
}

\def\hc#1{% heading C
 \twelverm
\vskip15pt
 #1\/
 \rm
}

\def\tx{% text
% \ifnum\LastMac=\Hae \else
%  \ifnum\LastMac=\Hbe \else
%   \ifnum\LastMac=\Skipe \else \One
%   \fi
%  \fi
% \fi
 \Fullout
 \twelvepoint\rm
}
\def\tf{% figure captions
 \Fullout
 \lineskip=12pt
 \tenpoint\rm
}

%\output{\OutputPage}

%\def\OutputPage{
% \shipout\vbox{\unvbox255}
%}
\nopagenumbers

\input psfig.tex

\headline={\twelverm 16th ECRS GR2 Lecture  \hfil  Page \folio}
\def\gr{$\gamma$-ray }
\def\grs{$\gamma$-rays }
\ct Diffuse gamma-ray emission: Galactic and extragalactic\par

\ca Martin Pohl\par

\aa DSRI, Juliane Maries Vej 30, 2100 Copenhagen \O, Denmark\par

\ha Abstract\par

\tx Here is reviewed our current understanding of Galactic and extragalactic
diffuse $\gamma$-ray emission. The spectrum of the extragalactic $\gamma$-ray background above 30 MeV can be well described by a power law
with photon index $\alpha$=2.1. In the 2-10 MeV range a preliminary analysis 
of COMPTEL data indicates a lower intensity than previously found, with
no evidence for an MeV bump. Most of the models of a truly diffuse
background seem to be in conflict with the
observed spectrum. Though AGN are the most likely input
from discrete sources, two independent attempts to model the high energy
background 
as the superposition of unresolved AGN indicate that AGN underproduce
the observed intensity. Therefore the origin of the extragalactic $\gamma$-ray background is still unknown.

The Galactic diffuse $\gamma$-ray continuum is more intense than expected both
at very low energies ($\le$ 100 keV) and at high energies ($\ge$ 1 GeV). 
The published models for these excesses all involve cosmic ray electron 
interactions. While the low energy excess may have something to do with
in-situ acceleration of electrons, the
excess at high energies may be understood if the sources of cosmic ray
electrons are discrete. The measured energy spectrum of the diffuse
Galactic \gr continuum radiation thus may provide new insights into the
acceleration of cosmic rays.

\ha The extragalactic gamma-ray background\par

\tx The extragalactic diffuse emission at \gr energies has interesting 
cosmological implications since the bulk of these photons suffer little or no 
attenuation during their propagation from the site of origin.
Before the launch of the Compton Gamma-Ray Observatory (CGRO),
several balloon experiments
(White et al. 1977; Sch\"onfelder et al. 1980) and the \gr spectrometer
flown aboard three Apollo flights (Trombka et al. 1977) showed the presence of
a feature in the few MeV range, that was in excess of the extrapolated
hard X-ray continuum. At higher energies, above 35 MeV, the SAS-2 satellite
provided the first clear evidence for the existence of an extragalactic 
\gr component (Fichtel et al. 1975).

The first all-sky survey in low energy \grs (1 MeV -- 30 MeV) has been
performed by COMPTEL and at higher energies, above 30 MeV, by EGRET on board
CGRO. The improved sensitivity, low
instrumental background and a large field of view of these instruments
have resulted in significantly improved measurements of the extragalactic
\gr background. In the following I shall briefly summarize the recent
analysis results, and then I shall discuss the implications of these
new findings on the origin of the extragalactic diffuse emission and 
current models thereof.

\hb COMPTEL results (1-30 MeV)\par

\tx In the 1-10 MeV band diffuse studies tend to be complicated by
difficulties in fully accounting for the instrumental background,
which is in general composed of `prompt' and `long-lived' components.

The prompt background is instantaneously produced by interactions of
energetic particles in the spacecraft and thus it modulates with
the instantaneous local cosmic-ray flux  which is monitored by the veto dome.
A linear extrapolation to zero veto count rate is used to eliminate the
prompt background contribution.

The long-lived background is caused by de-excitations of activated radioactive
isotopes with long half lives, for which the decay rate is not directly 
related to the instantaneous cosmic ray flux. The long-lived background events
are identified by their characteristic decay lines in the detector spectra.
Monte-Carlo simulations of the isotope decay are then used to determine
the absolute contribution of each of the isotopes.

\centerline{\psfig{figure=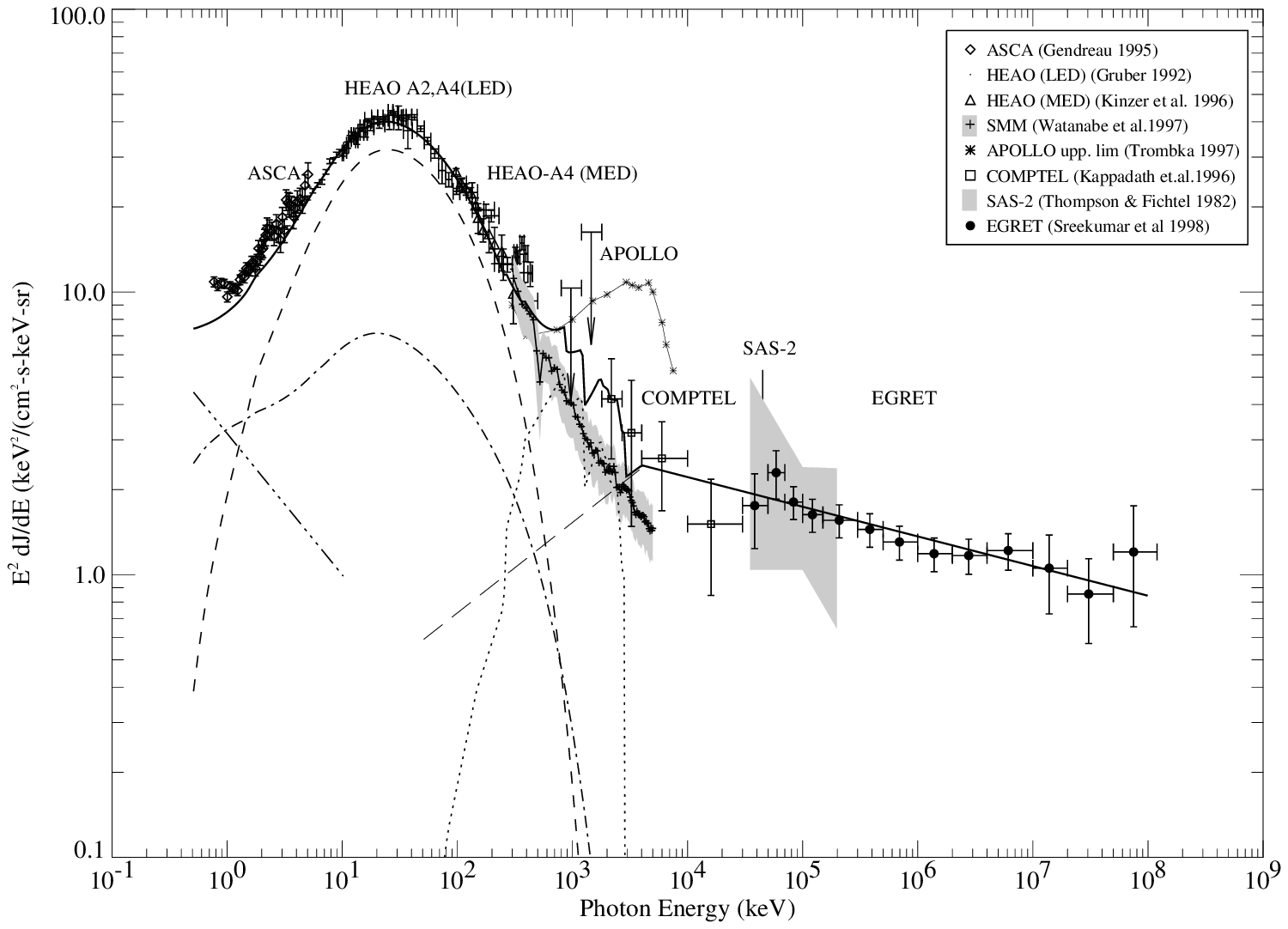,width=14.7cm,clip=}}

\noindent
{\tf Fig.1. Multiwavelength spectrum of the diffuse extragalactic
emission from X-rays to \grs taken from Sreekumar et al. (1997).
The estimated contributions from Seyfert I (dot-dashed) and Seyfert II (dashed)
are from the model by Zdiarski (1996); steep-spectrum quasar contribution
(triple dot-dashed) is taken from Chen, Fabian and Gendreau (1997);
Type Ia supernovae (dotted) is from The et al. (1993). Also included is
a possible
blazar contribution (long dashed) assuming an average power law index of
-1.7 below 4 MeV (McNaron-Brown et al. 1995) and -2.15 at higher energies
(Mukherjee et al. 1997). The thick solid line indicates the sum of all
components.} 

The diffuse flux measured by COMPTEL still includes contributions from the 
Galactic diffuse emission and \gr point sources in the field of view. 
The results below 9 MeV should, anyway, be considered preliminary. Nevertheless
it is clear to date that the 2-9 MeV flux is significantly lower than 
measured in the pre-COMPTEL era. There is no evidence for an MeV-bump
(Kappadath et al. 1996), a result supported also by a
recent analysis of SMM data
(Watanabe et al. 1997). The 9-30 MeV flux is compatible with earlier
measurements and also with the extrapolation of the EGRET spectrum.
The measured 9-30 MeV spectra from the Virgo and South Galactic Pole
regions are consistent with each other and hence with an isotropic nature
for the diffuse radiation (Kappadath et al. 1997).
The new results are shown in Figure 1 compared with the earlier
measurements.

\hb EGRET results (30 MeV to 100 GeV)\par

\tx At \gr energies above 10 MeV the analysis of diffuse emission
is complicated by difficulties in accurately accounting for the Galactic
diffuse emission. The following approach is adopted for the EGRET data.
The observed intensity $I_o$ is assumed to be made up of a
Galactic, $I_g$, and an extragalactic component, $I_{eg}$.
The extragalactic intensity is then derived by a straight line fit
to a plot of observed emission versus model prediction of Galactic emission
(Sreekumar et al. 1998). The spectrum is determined to be consistent with
a single power law of index ($2.1\pm 0.03$). The intensity above 100 MeV is
$(1.45\pm 0.05)\cdot 10^{-5}\,{\rm cm^{-2}sec^{-1} sr^{-1}}$, which may include 
some unaccounted extended Galactic diffuse emission. The spectrum
derived by Sreekumar et al. (1998) is also shown in Figure 1.

Earlier, independent analysis yielded a power law index ($2.11\pm 0.05$)
(Osborne, Wolfendale and Zhang 1994), or an index ($2.15\pm 0.06$)
and an integrated intensity above 100 MeV of
$(1.24\pm 0.06)\cdot 10^{-5}\,{\rm cm^{-2}sec^{-1} sr^{-1}}$
(Chen, Dwyer and Kaaret 1996), which are perfectly consistent 
with Sreekumar's result in the 
power law index, but indicate systematic uncertainties in the derivation
of the absolute intensity level. 

\hb Models of the extragalactic \gr background\par

\tx A large number of possible origins for the extragalactic diffuse
\gr emission have been proposed over the years. Theories of diffuse origin
include scenarios of baryonic symmetric universes (Stecker, Morgan, and 
Bredekamp 1971), primordial black hole evaporation (Page and Hawking 1976),
massive black holes that collapsed at redshifts of $z\sim 100$ (Gnedin and
Ostriker 1992), and annihilation of exotic particles (Silk and Srednicki
1984, Rudaz and Stecker 1991). All of the theories predict continuum and line
contributions that are not observed to date.

Models based on discrete source contributions have considered
a variety of source classes. The \gr intensity expected from normal galaxies 
has been estimated to be 5-10\% of what is observed
(e.g. Lichti, Bignami, and Paul 1978). Cosmic ray interactions
with intergalactic gas in groups and clusters of galaxies may add to this
(Dar and Shaviv 1995). However, the energy spectra of the Galactic and
intergalactic diffuse \grs are significantly different from those measured by
EGRET for the extragalactic diffuse radiation, hence these proposed sources
are unlikely to provide much more than 10\% of the observed extragalactic 
intensity.

It has been postulated for over 2 decades that unresolved
active galactic nuclei (AGN) might be the source of the extragalactic
diffuse emission (Bignami et al. 1979). Now the EGRET data prove that
a sub-class of AGN, namely blazars, are strong \gr emitters
(Mukherjee et al. 1997). A comparison of the spectra of \gr blazars with 
the diffuse spectrum gives ambiguous results. Averaging the best-fit indices
of power law
fits to the individual \gr spectra blazars yields ($2.15\pm 0.04$), in good
agreement with the index of the diffuse spectrum (Mukherjee et al. 1997).
Co-adding the individual intensity spectra of blazars on the other hand
results in an `average' spectrum which is distinctively softer than
the extragalactic background and not well represented by a power law
(Pohl et al. 1997a).

For any estimate of the intensity of diffuse emission from unresolved blazars, 
know\-ledge of the luminosity distribution and the evolution function is needed.
Chiang and Mukherjee (1998) have used the EGRET \gr blazar data alone to calculate the evolution and luminosity function of \gr loud AGN. These authors
report evidence for a low-luminosity cutoff in the \gr AGN luminosity function.
With this part of the luminosity function better constrained, they 
estimate that \gr loud AGN contribute an intensity of 
$I_{AGN} = (3\pm 1)\cdot 10^{-6}\ {\rm cm^{-2}sec^{-1} sr^{-1}}$
to the diffuse background,
which is about 25\% of the background observed by EGRET.

Several authors have estimated the contribution from blazars by assuming an
intimate relationship between \gr and radio emission from blazars.
Beyond applying the radio evolution function to the \gr emitting blazars,
the \gr fluxes have been compared with catalogued radio fluxes to derive
a luminosity correlation which can be integrated to obtain the
blazar contribution to the diffuse background (Padovani et al. 1993; Stecker, 
Salomon and Malkan 1993; Setti and Woltjer 1994; Erlykin and Wolfendale 1995). In most cases, the calculations are within 50\% of the observed intensity.
A word of caution about these estimates seems appropriate: there is certainly 
some sort of loose relation between radio and \gr emission of blazars
visible, for example, in that EGRET preferentially sees radio bright objects
(Mattox et al. 1997), but careful analysis shows that there is no
direct correlation with dispersion of less than a factor 2
(M\"ucke et al. 1997).

Recently, M\"ucke and Pohl (1997, 1998) have presented a calculation
in which they assumed that radio loud and \gr loud blazars are the same sources
in principle. They would share properties like evolution, geometry, and
energy input into the jet, but the actual radiation processes responsible for
radio and \gr emission would be different. Using a specific
inverse-Compton scattering model for the \gr production, these authors
deduce that AGN could provide only about 40\% of the observed diffuse 
background. However, they also state possible systematic problems in their
study in properly accounting for the contribution of BL Lacs.

All these studies either assume evolution similar to that observed
in radio source studies, or attempt to deduce the evolution directly from
the few high-z sources observed with EGRET. All of them need to introduce 
a redshift cutoff at which to stop integration of the blazar luminosity 
function.
The choice of redshift cutoff has considerable influence on the estimate 
of the diffuse radiation from  blazars. 

Gamma-rays of energy $\ge$ 10 GeV emitted at redshifts of $z\ge 4$ should
be reprocessed in a pair creation/annihilation
cascade before reaching the earth. One may therefore argue that the power law
shape up to 100 GeV exhibited by the observed diffuse background is
incompatible with a significant fraction of it originating from $z\ge 4$
objects.
Nevertheless, the systematic uncertainty imposed by the redshift cutoffs
is considerable. It is disturbing though, that two independent studies
indicate that AGN underproduce the diffuse background. If nothing else,
it tells us that to date we do not understand the origin of the
diffuse extragalactic \gr radiation.

\ha{The Galactic diffuse gamma-ray emission}

\tx Prior to the launch of CGRO, a number of observations of the Galactic
diffuse emission were made with instruments
covering generally non-overlapping energy ranges. At MeV energies the \gr spectrometer on SMM (Harris et al. 1990)
has provided high resolution spectra of the Galactic emission with
little spatial information. At higher energies above 50 MeV SAS-2
and COS-B have observed the Galactic plane with better spatial resolution, but
moderate energy resolution (for a review see Bloemen 1989). The intermediate 
energy range remained uncharted.

More recently the instruments OSSE, COMPTEL, and EGRET on CGRO have
provided us with a wealth of data of the Galactic diffuse \gr emission,
and thus with a much clearer understanding of the spectrum and the
spatial distribution of the line and continuum radiation. In the next
sections I will examine the main results of these observations, particularly
those which fail to conform with the standard picture. Finally I shall
discuss the constraints for cosmic ray physics thus derived.

\hb \gr line emission\par

\hc{Galactic $e^+$/$e^-$ annihilation line radiation}

\tx Since observations in the late 1970s gave
the first evidence of the 511 keV positron annihilation
line (Leventhal, MacCallum and Stang 1978), the Galactic center region 
has been observed by numerous experiments. These observations have 
not been able to determine the distribution of line emission due to limited
spatial resolution (Tueller 1993, Skibo, Ramaty and Leventhal 1992).
The situation has also been complicated by apparently time-variable
emission (Riegler et al. 1981, Leventhal et al. 1982), which
was thought to be 
caused by positrons escaping from compact sources (Ramaty et al. 1992).

Recently, the data taken with OSSE
have been combined with scanning observations
by TGRS and SMM to produce maps of the narrow Galactic 511 keV line emission
(Purcell et al. 1997). The resulting maps and modelling of the combined
data give evidence for three distinct spatial features: the Galactic plane, a
central bulge, and an extended emission region at positive latitudes
above the Galactic center. Purcell et al. find this asymmetric distribution
to be in good agreement with nearly all historic observations, without
invoking time variability. 
Considering fluxes rather than
the spatial distribution, supernovae seem
capable of producing positrons at the rate required to account for the
observed {511 keV} emission (Purcell et al. 1997).

The positive latitude
feature is suggestive of an outflow from the Galactic center. The extended
nature of the emission together with the lack of a high-density target 
seem to argue against jet activity from one or more of the black-hole
candidates
residing near the Galactic center. As an alternative it has been
proposed that the high-latitude feature is 
associated with a fountain of radioactive debris produced by enhanced
supernova activity in the Galactic center region (Dermer and Skibo 1997).

\hc{Galactic nuclear de-excitation \gr line emission}

\tx Nuclear de-excitation \gr lines provide a unique tracer for
low energy ($\sim 2-100$ MeV/nuc) cosmic ray nucleons. These are well
known from solar flares (Share, Murphy, and Ryan 1997),
but no compelling evidence was seen from other
sources prior to CGRO, although some claims were made.

An extensive evaluation of candidate \gr lines from nuclear interactions
was presented by Ramaty, Kozlovsky and Lingenfelter (1979).
The main candidate lines are from $^{12}C$ at 4.4 MeV and $^{16}O$ at 
6.1 MeV. The lines from energetic nuclei are broader than those
of ambient nuclei, hence both can be distinguished.
Gamma-ray spectroscopy of nuclear de-excitation lines thus provides a 
potentially powerful tool to study low energy cosmic ray nuclei and
their relative acceleration. It was not predicted, however, that 
CGRO would be able to detect such lines.

Preliminary results of COMPTEL observations of the inner Galaxy show
some, but not convincing evidence for line structure in the spectrum,
completely in line with theoretical expectations (Bloemen and Bykov 1997).  
It came as a surprise when early COMPTEL observations of the Orion
region revealed intense emission in the 3-7 MeV band which was soon attributed to Carbon and Oxygen de-excitation lines (Bloemen et al. 1994).
Later COMPTEL observations seemed to confirm the detection, though a
slightly different spatial distribution of the emission was obtained (Bloemen 
et al. 1997). OSSE has so far not detected this emission in Orion
(Murphy et al. 1996),
which can be reconciled with the COMPTEL results only if the source of
emission is very extended.

The existing observational limits on diffuse X-ray emission from
inverse brems\-strah\-lung, \gr continuum emission following $\pi^0$-decay,
and \gr line emission of heavier nuclei in the 1-3 MeV band,
require substantial fine tuning in attempts to model the Orion source as
Carbon and Oxygen de-excitation emission (Ramaty, Kozlovsky, and Tatischeff
1997, and references therein). However, alternative models have
not been able to explain both the observed spectrum and the
apparent lack of time 
variability (see Bloemen and Bykov 1997).

Recently it has been found by the COMPTEL team that background subtraction techniques used so far may be insufficient. A re-analysis of the
COMPTEL Orion data
is on the way and it is unclear to what extent the results will change.

\hb Galactic \gr continuum emission\par

\hc{Confusion and point sources}

\tx The analysis of the galactic diffuse emission can be seriously
complicated by unresolved galactic point sources which may have a sky 
distribution similar to that of interstellar
gas. Because of six objects already detected, 
pulsars are the most likely input from discrete sources. Many authors have
addressed this problem on the basis of pulsar emission models
(e.g. Yadigaroglu and Romani 1995; Sturner and Dermer 1996) and consistently estimated the contribution of pulsars to the diffuse \gr intensity, above
100 MeV integrated over the whole sky, to be a few percent.
Another strategy is to base the analysis only on
the observed properties of the six identified \gr pulsars, which also allows
an inspection of the spectrum of the unresolved pulsars (Pohl et al. 1997). 
It is found that pulsars contribute mostly at \gr 
energies above 1 GeV, and 
preferrentially exactly in the Galactic plane where they can provide more
than 20\% of the observed emission for a reasonable 
number of directly observable objects. 

Estimates for the contribution of discrete sources other than pulsars
are very uncertain due to the lack of clear identification of \gr
sources with any known population of Galactic objects. It is interesting to see
that roughly ten unidentified EGRET sources can be associated with
supernova remnants (SNR) or with OB associations, or with both (SNOBs)
(e.g. Sturner and Dermer 1995; Esposito et al. 1996; Yadigaroglu and Romani 
1997). Obviously these sources may also be radio-quiet
pulsars or highly dispersed radio pulsars.

\hc{The spatial distribution of \gr emission}

\tx Observations of the Magellanic Clouds with EGRET have finally settled
a long-standing debate on whether cosmic rays in the GeV energy range are 
Galactic or extragalactic. The 
\gr flux of the Large Magellanic Cloud is weakly less
(Sreekumar et al. 1992), and that of the Small
Magellanic Cloud is strongly less (Sreekumar et al. 1993) than expected, if 
cosmic ray protons were uniformly distributed in space.  
Therefore the bulk of
the locally observed protons at GeV energies must be Galactic, and we have
to 
think about which Galactic accelerators are capable of producing
cosmic rays with a source power of $\sim 10^{41}\ $erg/sec.

The spatial distribution of diffuse Galactic \grs is usually described
as `the gradient', that is a plot of the decline of \gr emissivity per H-atom
in the Galactic plane versus the galactocentric radius. This approach 
implicitely assumes that gas interactions (i.e. $\pi^0$ production and 
bremsstrahlung) dominate over inverse Compton scattering in the Galactic disk.
To investigate the \gr emission originating from $\pi^0$-decay and 
bremsstrahlung, we need some prior knowledge of the distribution 
of interstellar gas
in the Galaxy. This includes not only H$I$ but also H$_2$, which is 
indirectly traced by CO emission lines, and H$I\!I$, which is traced by
H$\alpha$ and pulsar dispersion measurements. Even in case of the directly 
observable atomic hydrogen we obtain only line-of-sight integrals,
albeit with some kinematic information. Any deconvolution of the velocity
shifts into distance is hampered by the line broadening of the
contribution from individual
gas clouds and by the proper motion of clouds with respect to the
main rotation flow. 

Different authors use different models of the 3D gas distribution in
the Galaxy and thus calculate different gradients (e.g. Strong and
Mattox 1996; Erlykin et al. 1996a). Detailed analysis of isolated gas clouds
in the solar vicinity shows that both the \gr emissivity 
and the CO line flux to molecular gas mass conversion factor, X,
can vary from place to place in the Galaxy
(Digel et al. 1995; Digel et al. 1996; Erlykin et al. 1996b).
Any comparison of gradients with the Galactic distribution of putative
cosmic ray sources should therefore be made with care. It may be safe to say,
however,
that the cosmic ray intensity decreases somewhat from the inner Galaxy
to the outer Galaxy.

\hc{The Galactic diffuse \gr spectrum at low energies}

\tx 
The OSSE (Purcell et al. 1996) and COMPTEL (Strong et al. 1994, 1996)
instruments have provided evidence that the diffuse Galactic continuum
emission extends down to photon energies below 100 keV, as shown in
Figure 2. In an analysis of
Galactic plane observations made with OSSE (Purcell et al. 1996), it was
found
that when the contribution from prominent point sources monitored
during simultaneous observations with SIGMA is subtracted from the 
Galactic center spectrum measured with OSSE, the residual
intensity is roughly
constant over the central radian of the Galaxy, but is lower by a factor 4
at $l\approx 95^\circ$ (Skibo et al. 1997). Estimates based on the
luminosities
and number-flux distributions of Galactic sources indicate that the point
source contribution to the hard X-ray emission from the Galactic plane is
less than 20\% (Yamasaki et al. 1997; Kaneda 1997).
The residual source-subtracted spectrum of this emission changes from a
photon index $\alpha =1.7$ at energies above 200 keV (Strong et al. 1994),
to a photon index $\alpha =2.7$ at lower energies (Purcell et al. 1996).
Thus the soft \gr continuum from the Galactic plane is more intense than 
the extrapolation of the higher energy emission. Observations of the
Galactic ridge in the hard X-ray range with GINGA (Yamasaki et al. 1997) and
RXTE (Valinia and Marshall 1998) indicate that the soft spectrum below 200 keV
extends down to about 10 keV energy, though the best spectral fit between 15 keV
and 150 keV gives a photon index of $\alpha =2.3$.

A hadronic origin for the hard X-ray/soft \gr continuum via inverse 
and secondary bremsstrahlung is excluded by the stringent observational
limits on the flux of nuclear \gr lines and $\pi^0$-decay \grs from
the inner Galaxy (Pohl 1998).  
Therefore the \gr continuum
emission in this energy band is most likely electron bremsstrahlung in
the interstellar medium. The power required in low energy ($<\,$10 MeV) cosmic 
ray electrons to produce a given amount of bremsstrahlung is a fixed
quantity that depends only on the energy spectrum of the radiating
electrons and weakly on the ionization state of the interstellar
medium. Attributing this power input to injection in cosmic ray electron
sources, it has been estimated that, integrated over the whole Galaxy, a
source power of about $4\times 10^{41}$ erg s$^{-1}$ (Skibo and
Ramaty 1993) or, if the bremsstrahlung emission extends down
to photon energies of 10 keV, up to $\sim 10^{43}$ erg sec$^{-1}$
(Skibo, Ramaty and Purcell 1996) in low energy

\centerline{\psfig{figure=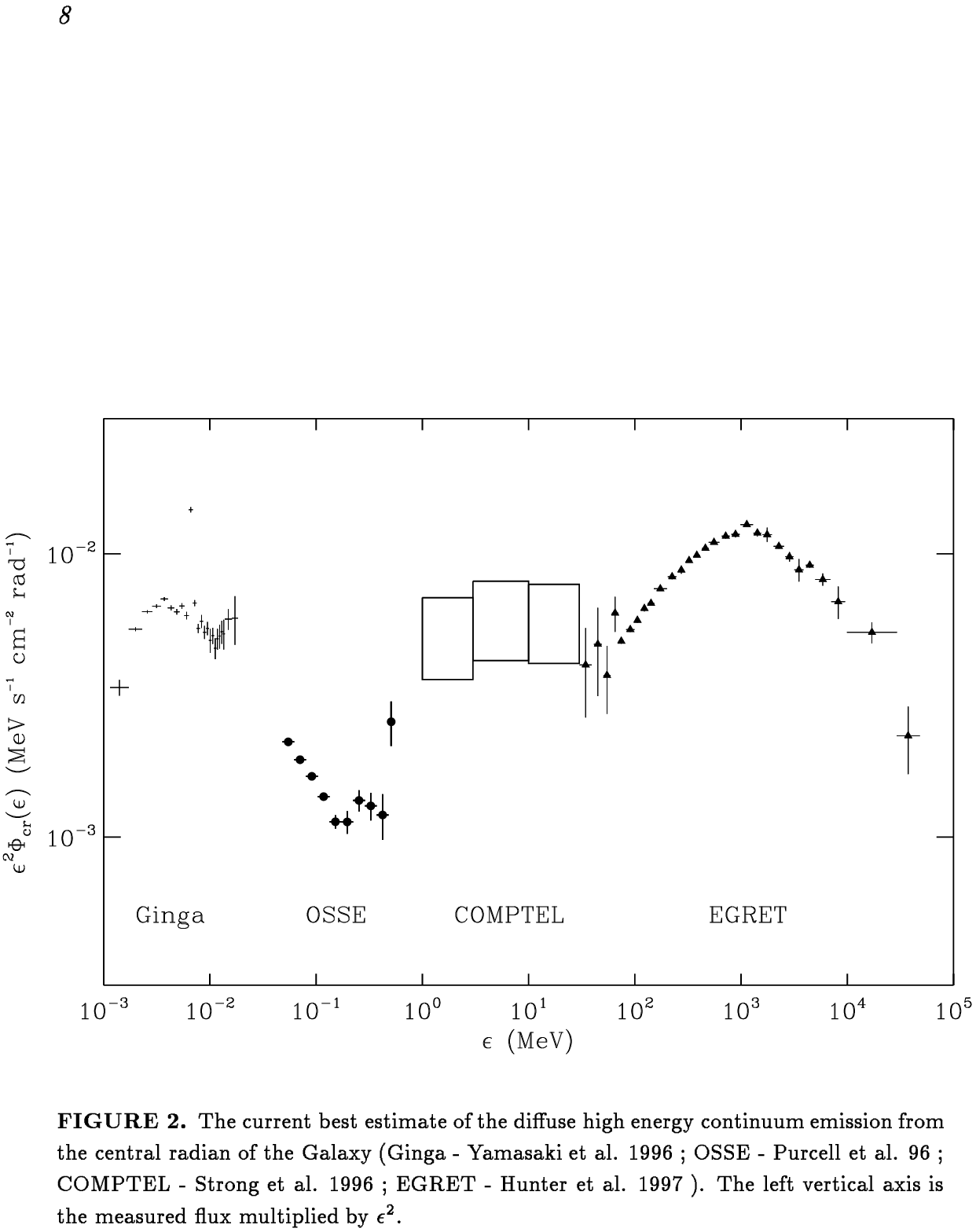,width=14.7cm,clip=}}

\noindent
{\tf Fig.2. The current best estimate of the diffuse high energy continuum
from the inner radian of the Galactic plane. The measured fluxes have been multiplied by $\epsilon^2$.}
\vskip12pt
\tx ($<$10
MeV) electrons is required, to retain sufficient electrons in the face of
severe Coulomb and ionization losses. This electron
power exceeds the power supplied to the nuclear cosmic ray component by
at least an order of magnitude. The energy losses of the required large population of
low energy electrons would be more than adequate to account
for the observed hydrogen ionization rate in the interstellar medium 
(Valinia and Marshall 1998). Proving the truly diffuse
nature of the galactic continuum emission below 1 MeV is of utmost
importance in pinning down the most relevant particle acceleration process
and to understand the interstellar medium ecosystem.

Recently, the extension of the bremsstrahlung continuum emission to
these low energies has been attributed to the existence of in-situ
stochastic electron acceleration by the interstellar plasma turbulence
(Schlickeiser 1997; Schlickeiser and Miller 1998), rather
than to the existence of a second electron source
component. This turbulence with a measured energy density of $\simeq
4\times 10^{-14}$ erg cm$^{-3}$ (Minter and Spangler 1997) is an important additional
energy source of cosmic ray particles.

\hc{The \gr spectrum at high energies}

\tx
The spatial and spectral
distributions of the diffuse emission within $10^\circ$ of the Galactic plane
have recently been compared with a model calculation of this emission
which is based on realistic interstellar
matter, photon distributions and dynamical balance (Hunter et al. 1997).
The distribution of the
total intensity above 100 MeV agrees surprisingly well with the model 
predictions. However, at higher energies, above 1 GeV, the model systematically
underpredicts the \gr intensity. If the model is scaled up by a factor 1.6, 
the model prediction and the observed intensity above 1 GeV agree well.
This deficit can be explained neither by a possible
miscalibration of EGRET, nor by spectral changes in the nucleonic $\pi^0$-decay
emission component (Mori et al. 1997), nor by
unresolved point sources like pulsars (Pohl et al. 1997b).

The diffuse model deficit above 1 GeV is visible also at higher latitudes,
e.g. in the plots of observed intensity versus Galactic diffuse model
shown in the paper of Sreekumar et al. (1998). Uncritical use of the nominal
Galactic diffuse model may therefore lead to apparent \gr excesses at higher
latitudes, which then may be mistaken as evidence for a \gr halo of exotic
origin.

Thus the
model displays a deficit of $\sim$40\% of the total observed emission
which depends, if at all, only weakly on location.
One feature of the models is the relatively soft
electron injection spectral index of s=2.4 (Skibo 1993), which is required
to account for the local electron spectrum above 50 GeV.
Consequently at energies above 1 GeV,
around 90\% of the model intensity is due to $\pi^0$-decay (i.e. hadronic
processes) and only 10\% is due to interactions of electrons.

The recent detections of non-thermal X-ray synchrotron radiation
from the four supernova remnants SN1006 (Koyama et al. 1995), RX J1713.7-3946
(Koyama et al. 1997), IC443 (Keohane et al. 1997), and Cas A (Allen et al. 1997),
and the subsequent detection of SN1006 at TeV energies (Tanimori et al. 1998)
and flux levels according to theoretical predictions (Pohl 1996),
support the hypothesis that Galactic cosmic ray electrons are accelerated
predominantly in SNR. It has been shown that, if this is
indeed the case, the local electron spectra above 30 GeV are
variable on time scales of about $10^5$ years (Pohl and Esposito 1998).
This variability 
stems from the Poisson fluctuations in the number of SNR in the solar
vicinity within a certain time period. While the electron spectra below 10
GeV are stable, the level of fluctuation increases with electron energy,
and above 100 GeV the local electron flux is more or less unpredictable.

Considering this time variability, an electron injection
index of s=2.0 is consistent with direct particle 
measurements if SNR are the dominant source of cosmic ray electrons.
While being entirely consistent with the local electron flux,
and with the radio synchrotron spectrum towards
the North Galactic Pole, the leptonic contribution to the diffuse Galactic
\gr emission above {1 GeV}
in the Galactic plane would increase to 30-48\% of the total observed intensity
for an injection index of s=2.0, depending on the assumed
spatial distribution
of SNR and on whether some dispersion of injection spectral 
indices is allowed (Pohl and Esposito 1998).
An electron injection index of s=2.0 may therefore
explain the bulk of the observed \gr excess over that predicted by the
Hunter et al. model.

\ha{References}

\tx 

G.E. Allen et al. {\twelveit Ap.J.} 487 (1997) L97-100

G.F. Bignami et al. {\twelveit Ap.J.} 232 (1979) 649-658

J.B.G.M. Bloemen. {\twelveit A.R.A.$\&$A.} 27 (1989) 469-517

H. Bloemen, A.M. Bykov. {\twelveit Proceedings of the 4th Compton Symposium}. 
AIP Conference Proceedings. 410 (1997) 249-268

H. Bloemen et al. {\twelveit A.$\&$A.} 284 (1994) L5-8

H. Bloemen et al. {\twelveit Ap.J.} 475 (1997) L25-28

A. Chen, J. Dwyer, P. Kaaret. {\twelveit Ap.J.} 463 (1996) 169-180

L.W. Chen, A.C. Fabian, K.C. Gendreau. {\twelveit M.N.R.A.S.} 285 (1997) 449-471

A. Dar, N. Shaviv. {\twelveit P.R.L.} 75 (1995) 3052-3055

C.D. Dermer, J.G. Skibo. {\twelveit Ap.J.} 487 (1997) L57-60

S.W. Digel et al. {\twelveit Ap.J.} 463 (1996) 608-622

S.W. Digel et al. {\twelveit Ap.J.} 441 (1995) 270-280

A.D. Erlykin et al. {\twelveit A.$\&$A.S.} 120 (1996a) C397-401

A.D. Erlykin et al. {\twelveit A.$\&$A.S.} 120 (1996b) C415-418

A.D. Erlykin, A.W. Wolfendale. {\twelveit J.Phys. G} 21 (1995) 1149-1165

J. Esposito et al. {\twelveit Ap.J.} 461 (1996) 820-827

C.E. Fichtel et al. {\twelveit Ap.J.} 198 (1975) 163-182

N.Y. Gnedin, J.P. Ostriker. {\twelveit Ap.J.} 400 (1992) 1-20

M.J. Harris et al. {\twelveit Ap.J.} 362 (1990) 135-146

S.D. Hunter et al. {\twelveit Ap.J.} 481 (1997) 205-240

H. Kaneda. {\twelveit PhD Dissertation}. University of Tokyo (1997)

S.C. Kappadath et al. {\twelveit A.$\&$A.S.} 120 (1996) C619-622

S.C. Kappadath S.C. et al. {\twelveit Proceedings of the 4th Compton Symposium}. AIP Conference Proceedings. 410 (1997) 1218-1221

J.W. Keohane et al. {\twelveit Ap.J.} 484 (1997) 350-359

K. Koyama et al. {\twelveit Nature} 378 (1995) 255-258

K. Koyama et al. {\twelveit P.A.S.J.} 49 (1997) L7-11

M. Leventhal, C.J. MacCallum, P.D. Stang. {\twelveit Ap.J.} 225 (1978) L11-14

M. Leventhal et al. {\twelveit Ap.J.} 260 (1982) L1-4

G.G. Lichti, G.F. Bignami, J.A. Paul. {\twelveit Ap.$\&$S.S.} 56 (1978) 403-414

J.R. Mattox et al. {\twelveit Ap.J.} 481 (1997) 95-115

K. McNaron-Brown et al. {\twelveit Ap.J.} 451 (1995) 575-584

A.H. Minter, S.R. Spangler. {\twelveit Ap.J.} 485 (1997) 182-194

M. Mori. {\twelveit Ap.J.} 478 (1997) 225-232

R. Mukherjee et al. {\twelveit Ap.J.} 490 (1997) 116-135

R.J. Murphy et al. {\twelveit Ap.J.} 473 (1996) 990-997

A. M\"ucke et al. {\twelveit A.$\&$A.} 320 (1997) 33-40

A. M\"ucke, M. Pohl. {\twelveit Proceedings of the 4th Compton Symposium}. 
AIP Conference Proceedings. 410 (1997) 1233-1236

A. M\"ucke, M. Pohl. {\twelveit M.N.R.A.S.} (1998) submitted

J.L. Osborne, A.W. Wolfendale, J. Zhang. {\twelveit J.Phys. G} 20 (1994) 1089-1101

P. Padovani et al. {\twelveit M.N.R.A.S.} 260 (1993) L21-24

D.N. Page, S.W. Hawking. {\twelveit Ap.J.} 206 (1976) 1-7

M. Pohl. {\twelveit A.$\&$A.} (1998) submitted

M. Pohl. {\twelveit A.$\&$A.} 307 (1996) L57-59

M. Pohl et al. {\twelveit A.$\&$A.} 326 (1997a) 51-58

M. Pohl et al. {\twelveit Ap.J.} 491 (1997b) 159-164

M. Pohl, J. Esposito. {\twelveit Ap.J.} 507 (1998) in press

W.R. Purcell et al. {\twelveit Ap.J.} 491 (1997) 725-748

W.R. Purcell et al. {\twelveit A.$\&$A.S.} 120 (1996) C389-392

R. Ramaty et al. {\twelveit Ap.J.} 392 (1992) L63-66

R. Ramaty, B. Kozlovsky, R.E. Lingenfelter. {\twelveit Ap.J.S.} 40 (1979) 487-526

G.R. Riegler et al. {\twelveit Ap.J.} 248 (1981) L13-16
 
S. Rudaz, F.W. Stecker. {\twelveit Ap.J.} 368 (1991) 406-410

R. Schlickeiser. {\twelveit A.$\&$A.} 319 (1997) L5-8

R. Schlickeiser, J.A. Miller. {\twelveit Ap.J.} 492 (1998) 352-378

V. Sch\"onfelder, F. graml, F.-P. Penningsfeld {\twelveit Ap.J.} 240 (1980) 350-362

G. Setti, L. Woltjer. {\twelveit Ap.J.S.} 92 (1994) 629-632

G.H. Share, R.J. Murphy, J. Ryan. {\twelveit Proceedings of the 
4th Compton Symposium}. AIP Conference Proceedings. 410 (1997) 17-36

J. Silk, M. Srednicki. {\twelveit P.R.L.} 53 (1984) 624-627

J.G. Skibo et al. {\twelveit Ap.J.} 483 (1997) L95-98

J.G. Skibo. {\twelveit PhD Dissertation} University of Maryland (1993)

J.G. Skibo, R. Ramaty. {\twelveit A.$\&$A.S.} 97 (1993) 145-148

J.G. Skibo, R. Ramaty, W.R. Purcell. {\twelveit A.$\&$A.S.} 120 (1996) C403-406

J.G. Skibo, R. Ramaty, M. Leventhal. {\twelveit Ap.J.} 397 (1992) 135-147

P. Sreekumar et al. {\twelveit Ap.J.} 494 (1998) 523-534

P. Sreekumar et al. {\twelveit P.R.L.} 70 (1993) 127-129

P. Sreekumar et al. {\twelveit Ap.J.} 400 (1992) L67-70

F.W. Stecker, D.L. Morgan, J. Bredekamp. {\twelveit P.R.L.} 27 (1971) 1469-1472

F.W. Stecker, M.H. Salamon, M. Malkan. {\twelveit Ap.J.} 410 (1993) L71-74

A.W.Strong, J.R. Mattox. {\twelveit A.$\&$A.} 308 (1996) L21-25

A.W.Strong et al. {\twelveit A.$\&$A.} 292 (1994) 82-91

A.W.Strong et al. {\twelveit A.$\&$A.S.} 120 (1996) C381-388

S.J. Sturner, C.D. Dermer. {\twelveit Ap.J.} 461 (1996) 872-883

S.J. Sturner, C.D. Dermer. {\twelveit A.$\&$A.} 293 (1995) L17-20

T. Tanimori et al. {\twelveit Ap.J.} 497 (1998) L25-28

L.-S. The, M.D. Leising, D.D. Clayton. {\twelveit Ap.J.} 403 (1993) 32-36

J.I. Trombka et al. {\twelveit Ap.J.} 212 (1977) 925-936

J. Tueller. {\twelveit Proc. Compton Symp.}, Eds. Friedlander and
Gehrels, New York, AIP 280 (1993) 97-100

A. Valinia, F.E. Marshall. {\twelveit Ap.J.} (1998) in press, astro-ph/9804012

K. Watanabe et al. {\twelveit Proceedings of the 4th Compton Symposium}. 
AIP Conference Proceedings. 410 (1997) 1223-1226

R.S. White et al. {\twelveit Ap.J.} 218 (1977) 920-927

I.-A. Yadigaroglu, R.W. Romani. {\twelveit Ap.J.} 476 (1997) 347-356

I.-A. Yadigaroglu, R.W. Romani. {\twelveit Ap.J.} 449 (1995) 211-215

N.Y. Yamasaki et al. {\twelveit A.$\&$A.S.} 120 (1996) C393-396

N.Y. Yamasaki et al. {\twelveit Ap.J.} 481 (1997) 821-831

A.A. Zdiarski. {\twelveit M.N.R.A.S.} 281 (1996) L9-12

\vfill\eject
\end\bye